\documentclass[twocolumn]{aastex62}
\usepackage{amsmath}
\usepackage{graphicx}
\usepackage{color}

\usepackage{amsmath}
\usepackage{graphicx}

\newcommand{\beq}{\begin{equation}}
\newcommand{\eeq}{\end{equation}}

\begin{document}

\title{Topological Data Analysis of Black Hole Images}

\author{Pierre Christian}
\affiliation{Physics Department, Fairfield University, 1073 N Benson Rd, Fairfield, CT 06824}
\author{Chi-kwan Chan}
\affiliation{Department of Astronomy, University of Arizona, 933 North Cherry Avenue, Tucson, AZ 85721, USA}
\affiliation{Data Science Institute, University of Arizona, 1230 N. Cherry Ave., Tucson, AZ 85721}
\affiliation{Program in Applied Mathematics, University of Arizona, 617 N. Santa Rita, Tucson, AZ 85721}
\author{Anthony Hsu}
\affiliation{Department of Computer Science, University of Arizona, 1040 4th St, Tucson, AZ 85721, USA}
\author{Feryal \"Ozel}
\author{Dimitrios Psaltis}
\affiliation{Department of Astronomy, University of Arizona, 933 North Cherry Avenue, Tucson, AZ 85721, USA}
\author{Iniyan Natarajan}
\affiliation{Wits Centre for Astrophysics, University of the Witwatersrand, 1 Jan Smuts Avenue, Braamfontein, Johannesburg 2050, South Africa}
\affiliation{South African Radio Astronomy Observatory, Observatory 7925, Cape Town, South Africa}

\begin{abstract}
Features such as photon rings, jets, or hot spots can leave particular topological signatures in a black hole image. As such, topological data analysis can be used to characterize images resulting from high resolution observations (synthetic or real) of black holes in the electromagnetic sector. We demonstrate that persistent homology allows for this characterization to be made automatically by counting the number of connected components and one-dimensional holes. Further, persistent homology also allows for the distance between connected components or diameter of holes to be extracted from the image. In order to apply persistent homology on synthetic black hole images, we also introduce metronization, a new algorithm to prepare black hole images into a form that is suitable for topological analysis.\\
\end{abstract}

\section{Introduction}

The Event Horizon Telescope (EHT) has recently showcased the feasibility of ground based very-long-baseline-interferometers (VLBIs) to produce images of supermassive black holes \citep{PaperI,PaperII,PaperIII,PaperIV,PaperV,PaperVI}. In performing analysis for its dataset, the EHT collaboration utilizes a suite of synthetic observations generated by ray-tracing General Relativistic Magnetohydrodynamic (GRMHD) simulations assuming certain radiative transfer models \citep{PaperV}.

As the EHT project expands, the variety of parameters included in the synthetic image calculations necessarily increases. Modifications to the synthetic image calculations can vary from simply increasing the range and resolution of previous calculations (e.g., computing more inclination angles), to changing the underlying astrophysical model (e.g., utilizing different radiative transfer or GRMHD models), to changing the fundamental physical theories (e.g., utilizing non general-relativistic theories).

This bounty of synthetic images presents a problem. Given their quantity, it will be prohibitively expensive to compare each synthetic image with the observed dataset. Thus, a method needs to be devised to quickly classify the synthetic images based on some broad features, which can then be quickly compared with either the observed dataset or some prior knowledge of the black hole in question. 

Further, interferometric imaging such as the one conducted by the EHT relies on reconstructing real-space images from incomplete observations of the target in Fourier space \citep{Honma, Akiyama, Bouman, Chael}. Different reconstruction methods, as well as different choices of parameters within one reconstruction method, can produce widely differing images. Given the large amount of image-domain reconstructions that are produced in projects such as the EHT, it is also important to be able to quickly classify these images into bins based on some broad features. 

Apparent in the first EHT observations \citep{PaperI, PaperVI}, and in the synthetic observations of the M87 black hole \citep{PaperV} is a particular topological signature of the black hole image: a bright ring surrounding a dim region. We propose that topological signatures such as the existence of such a ring is a good set of features to classify black hole images. 

Topological data analysis (TDA) is a set of techniques that allows for robust analysis of the topological characteristics of a dataset to be made. In this work, we introduce a method to classify black hole images (synthetic or real) by the application of persistent homology, a TDA technique that measures certain topological invariants of a dataset. In persistent homology, one first converts an image into a topological space associated with the datapoints of the image. Once in the form of a topological space, standard mathematical techniques can be used to compute the topological properties of the image. More details on TDA and persistent homology can be found in \cite{GCarlssonTopologyandData} or in the book \cite{EandH}.

Persistent homology converts a high dimensional dataset (a black hole image) into more manageable low dimensional information (its topological characteristics), and is thus an example of a dimensionality reduction technique. In the study of black hole shadows, dimensionality reduction in the form of principal component analysis (PCA) has previously been utilized to decompose synthetic black hole images \citep{Lia1} and theoretical models of black hole shadows \citep{Lia2} into their eigencomponents. In contrast to PCA, where the transformation from high to low dimension spaces is linear, persistent homology applies a non-linear transformation. This nonlinearity allows persistent homology to retain information that is lost by linear dimensionality reduction techniques.

In order to apply TDA to the large image set, we develop a two-step algorithm. The first step, which we call \emph{metronization}, is a dimensional reduction procedure that condenses a black hole image into simple graphs that retain all of the topological information of the image. This step both transforms a black hole image into a form amenable to TDA as well as greatly accelerating the TDA calculations. The second step is to run persistent homology algorithms on the metronization output. Currently, there are many available open-source software to perform the later computational topology tasks. In this project, we utilized the open-source code Dionysus 2\footnote{https://mrzv.org/software/dionysus2/} \citep{Dionysus2}. Other examples of computational topology software include Ripser++\footnote{https://github.com/simonzhang00/ripser-plusplus} \citep{ripserplusplus} which possesses graphic processing unit (GPU) acceleration and the Topology Toolkit\footnote{https://topology-tool-kit.github.io/} \citep{TTK}. 

This paper is organized as follows: in section \S \ref{s2} we give an introduction to persistent homology; in section \S \ref{s3} we describe metronization, our pre-processing technique to convert black hole images into a form suitable for persistent homology; in section \S \ref{s4} we show examples of persistent homology applied to mock black hole images; and in section \S \ref{s5} we supply our concluding remarks.

\section{Persistent Homology} \label{s2}

An abstract simplicial complex can be thought of as a collection of $n$-dimensional triangles, i.e., a collection of 0-dimensional points, 1-dimensional lines, 2-dimensional triangles, 3-dimensional tetrahedrons, etc \citep[see e.g.,][for a rigorous definition]{LeeTopology}. Given such a structure, one can compute the topological invariants associated with the homology groups -- a set of mathematical objects that are closely related to the number of holes on a space. Considering our datapoints as points in $\mathbb{R}^2$ equipped with its standard metric, 
\begin{equation}
d = \sqrt{x^2+y^2} \; ,    
\end{equation}
where $(x,y)$ are coordinates on $\mathbb{R}^2$ and $d$ the distance on the image plane, we can transform our datapoints into a simplicial complex using the Vietoris-Rips algorithm \citep{Vietoris}. The algorithm is as follows: for a length-scale $l$,
\begin{enumerate}
    \item consider each data point as a 0-simplex
    \item two points within a distance $l$ are connected with a 1-simplex
    \item three points that are connected to each other by a 1-simplex form a filled triangle (a 2-simplex)
\end{enumerate}
This algorithm can then be repeated for many $l$'s, and thus gives the topological invariants as a function of $l$. 

For our purposes, the most relevant topological invariants are the Betti numbers, which can be computed from a simplicial complex in a straightforward manner: for a space $X$, they are defined as
\begin{equation}
    b_n(X) = \mathrm{Rank}[H_n(X)] \; ,
\end{equation}
where $\mathrm{Rank}[G]$ is the rank of a group $G$ and $H_n(X)$ is the $n$th homology group of $X$, which is defined as,
\begin{equation}
    H_n = \mathrm{Ker}(\partial_{n})/\mathrm{Im}(\partial_{n+1})\; ,
\end{equation}
where $\partial_n$ is the $n$th dimensional boundary operator of the simplicial complex and (as from now on) the dependence on the space $X$ has been suppressed. Equivalently, the Betti number can be computed by
\begin{equation}
    b_n= \mathrm{dim}[\mathrm{Ker}(\partial_{n})] - \mathrm{dim}[\mathrm{Im}(\partial_{n+1})]\; ,
\end{equation}
where $\mathrm{dim}[\;]$ is the dimension operator. 

As black hole images are two-dimensional and $b_n$ is zero for $n$ greater than the dimension of the space, we only need to consider $b_0$, which counts the number of connected components, and $b_1$, which counts the number of one-dimensional holes (i.e., circles). For example, a space with two holes will have
\begin{equation}
    H_2 = \mathbb{Z} \oplus \mathbb{Z} \; ,
\end{equation}
which gives $b_2(U)=2$. 

\subsection{Persistent homology in practice}

\begin{figure}[t]
    \centering
    \includegraphics[width=\linewidth]{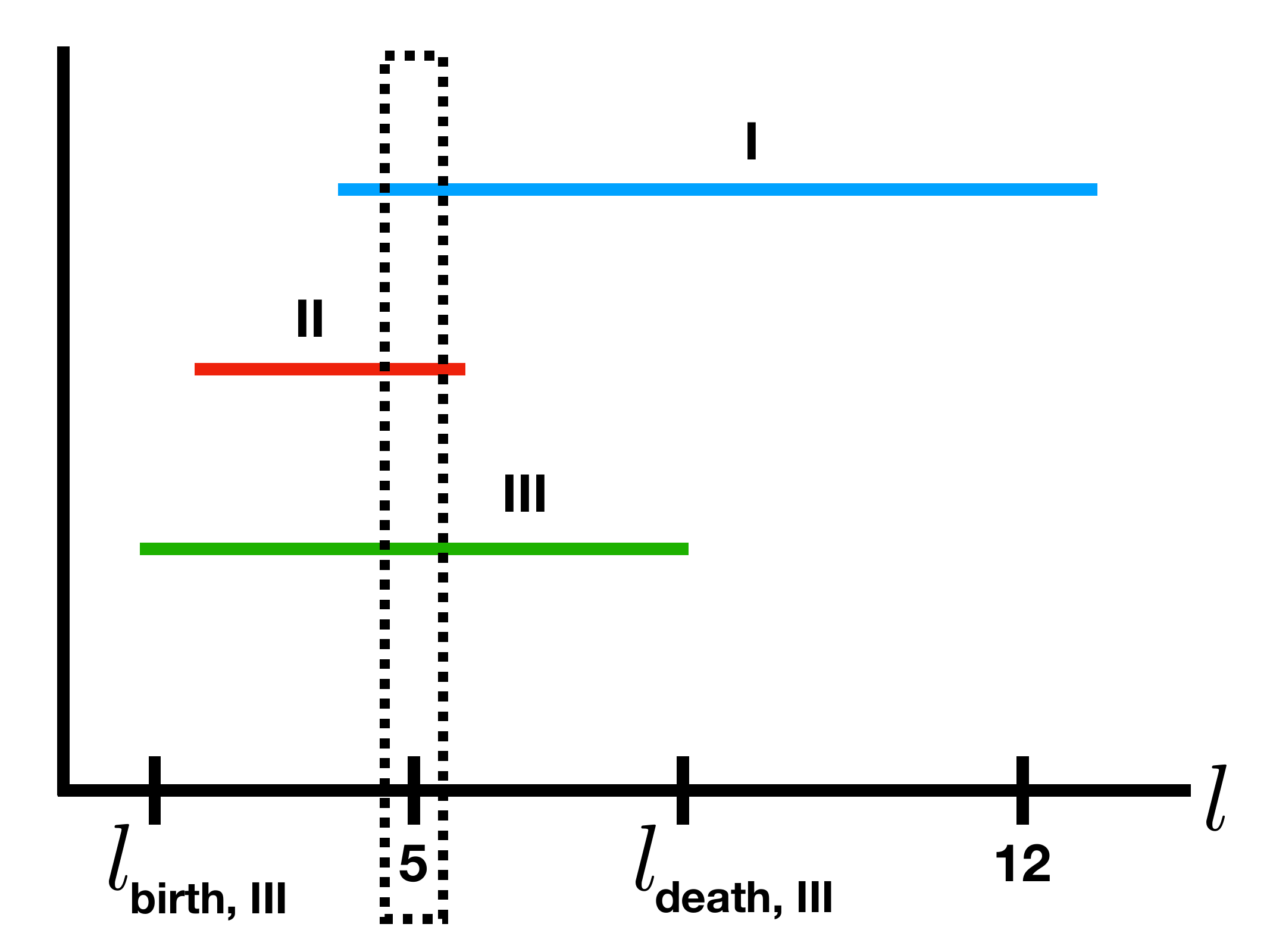} \\
    \caption{A schematic example of a barcode diagram.}
    \label{fig:barcode}
\end{figure}

\begin{figure}[t]
    \centering
    \includegraphics[width=\linewidth]{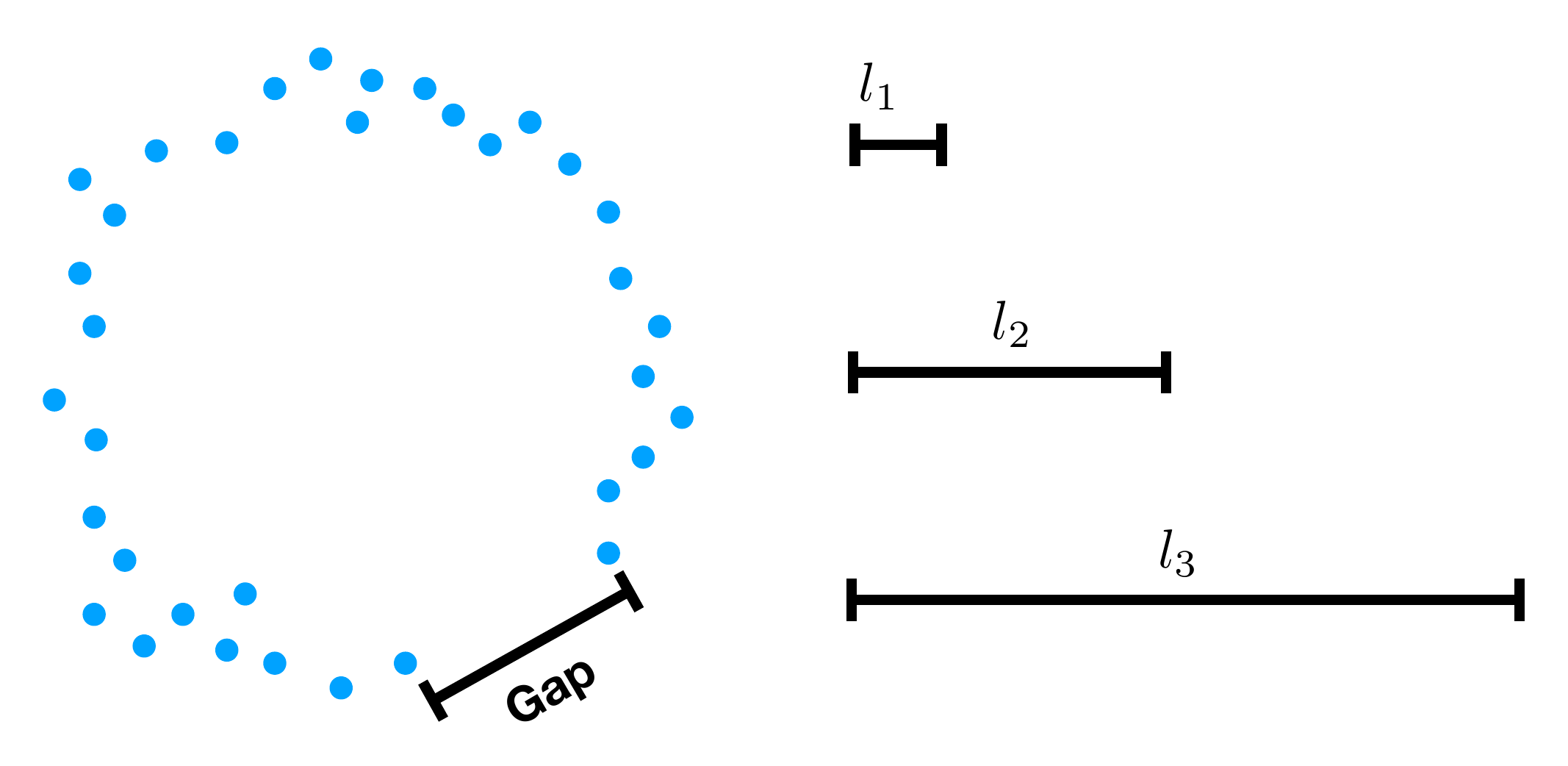} \\
    \caption{Birth and death of topological features: due to the gap in the data, $l_1$ is too short to connect all the datapoints to form a circular object and so $b_1(l_1)=0$. In contrast, $l_2$ is long enough that the datapoints are connected together into a disk-like object with a hole in the middle, giving $b_1(l_2)=1$. However, as $l$ is increased, this hole eventually gets filled in, and is no longer detected when $l=l_3$, i.e., $b_1(l_3)=0$.}
    \label{fig:birthdeathschem}
\end{figure}

A barcode diagram is a typical output of topological data analysis that encodes the Betti numbers found by the persistent homology algorithm as a function of $l$. A schematic example of a barcode diagram is shown in Figure \ref{fig:barcode}. 

Suppose that the diagram in Figure \ref{fig:barcode} is the $b_1$ barcode diagram (one barcode diagram can be plotted for each Betti number). The horizontal axis of a barcode diagram is $l$, the characteristic length-scale. At a particular $l$, a manifold is formed out of the data points, the number of one-dimensional holes in this manifold is computed, and each hole is plotted as a point along the vertical axis at said $l$ (at no particular order in the horizontal axis). Figure \ref{fig:barcode} then shows that when $l=5$ there are three holes in the image, as there are three points plotted at $l=5$. Similarly, when $l=12$, there is only one hole in the image. 

As $l$ is increased, holes are \emph{born} when $l$ becomes large enough to connect the datapoints forming the boundary of the hole. However, these holes eventually \emph{die} when $l$ becomes large enough that they get filled (see Figure \ref{fig:birthdeathschem}). As such, for every detected hole, the barcode diagram for $b_1$ starts at the $l$ where the hole is first detected and ends at the $l$ where the hole gets filled in. The barcode diagram for $b_0$ can be analogously described, with the number of connected components taking the place of the number of holes.

Note that unlike homology, persistent homology is sensitive to distances and sizes. The characteristic length where the barcode diagram ends, $l_{\textrm{death}}$, for the $b_1$ diagram is a measure of the size of the hole. Particularly, for a dataset with a circular hole of radius $R$, 
\begin{equation}
    l_{\textrm{death}} = 2R\cos{\pi/6} \approx 1.73 R \; ,
\end{equation}
and for a dataset consisting of two separate clumps of datapoint, $l_{\textrm{death}}$ for the $b_0$ diagram is a measure of the distance between these two clumps.

Most topological signals are short-lived, i.e., with a "lifespan" of
\begin{equation}
    \Delta l = l_{\textrm{death}} - l_{\textrm{birth}}
\end{equation}
being small. For our purposes, these short-lived topological signatures are not interesting, as they are typically the result of noise in the data spuriously forming topological objects such as small rings. As such, we define a cutoff $L$ on the lifespan, where if
\begin{equation}
    \Delta l > L \; ,
\end{equation}
we consider the topological signal to be robust or \emph{persistent}. In the $b_1$ diagram, a cutoff of $L$ means that only circles with radius at least $\approx 1.73 L$ will pass our cutoff. For $b_1$ diagrams, we also impose a lower cutoff: any holes that are born at $l=0$ must be due to the noise as at $l=0$ the datapoints should all be disconnected.

\begin{figure*}[t]
    \centering
    \includegraphics[width=\textwidth]{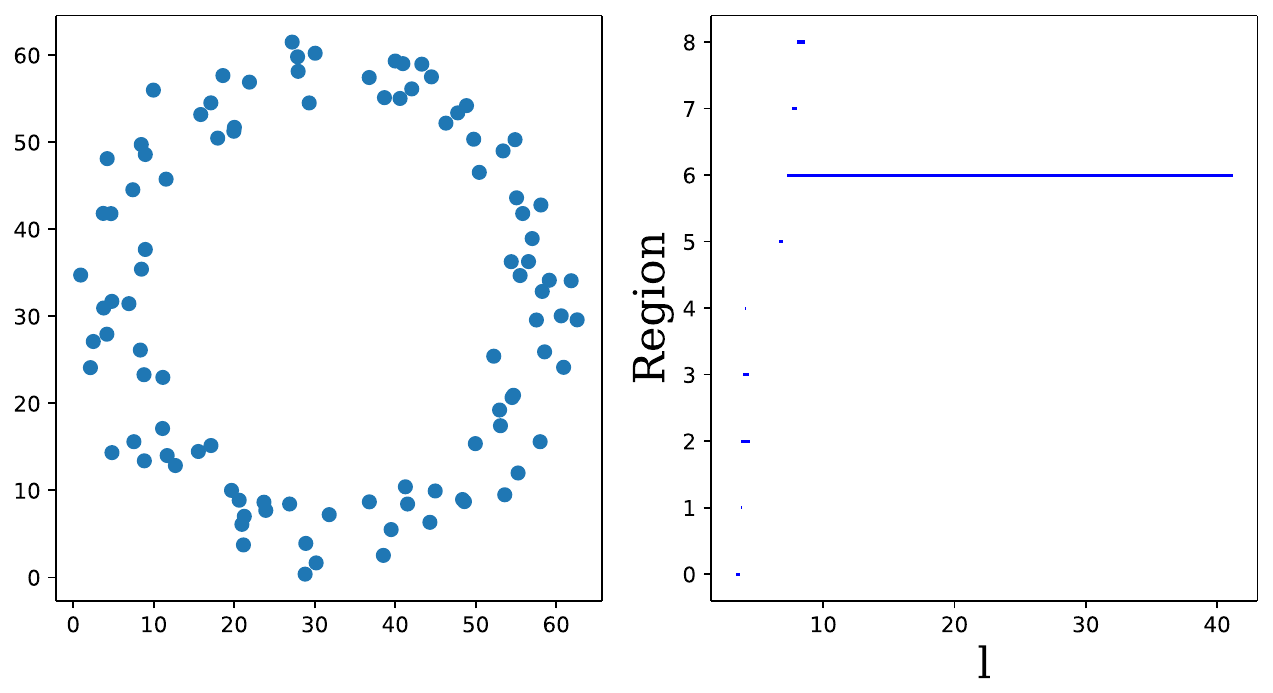} 
    \caption{Data points in a ring (left) of diameter $\sim0.1$, along with its $b_1$ barcode diagram (right). The barcode diagrams show the detection of one persistent hole. The $b_1$ barcode terminates at $l=2R\cos{(\pi/6)}$, as expected for an equilateral triangle inscribed in the circular hole.}
    \label{fig:example1}
\end{figure*}

\begin{figure*}[t]
    \centering
    \includegraphics[width=\textwidth]{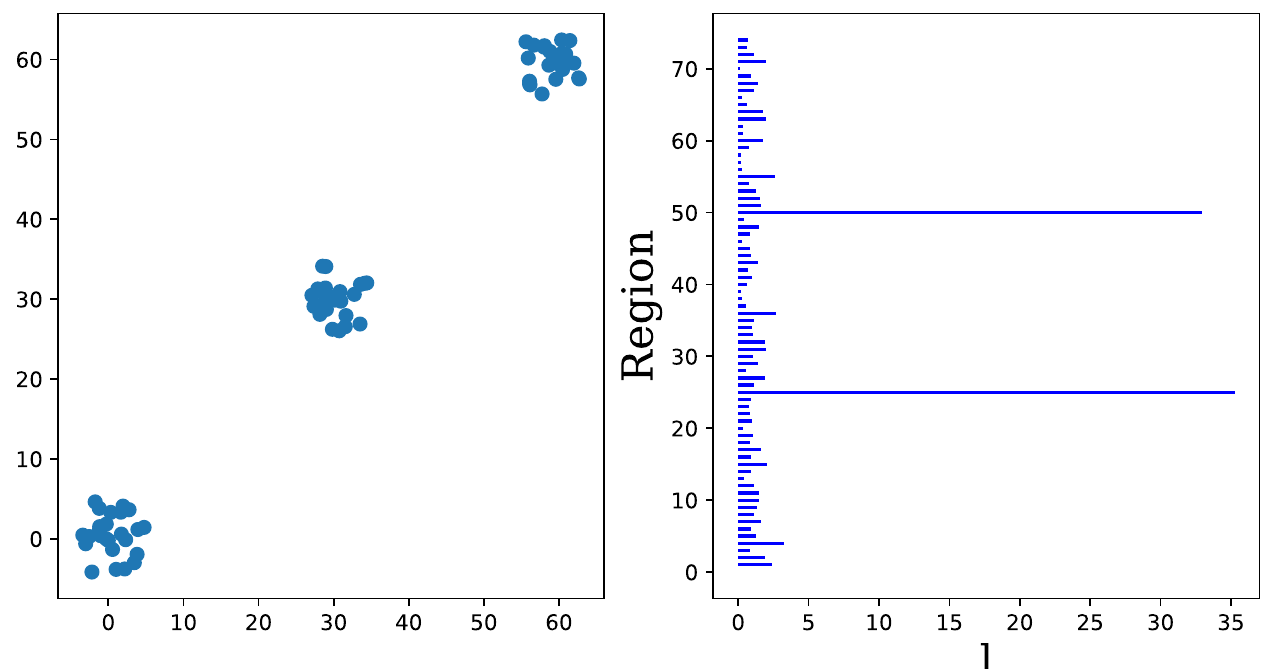} 
    \caption{Data points in three distinct connected components (top) and their associated $b_0$ barcode diagram (bottom). Note that the first connected component that trivially exists for any non-empty dataset (with $l_{\textrm{birth}}=0$ and $l_{\textrm{death}} \to \infty$) is not plotted.
    The barcode diagram detects three persistent connected components and displays the distances between each connected components in the $l_{\textrm{death}}$'s.}
    \label{fig:example3}
\end{figure*}

Figure \ref{fig:example1} shows a dataset consisting of datapoints arranged in a noisy circle (by noisy we mean the dataset contain points that deviate from a perfect circle) and the barcode diagram corresponding to $b_1$. As the barcode diagram attests, there is one component that is much longer than the rest. Figure \ref{fig:example3} shows the $b_0$ diagram for a dataset consisting of three data clouds.

\section{Metronization} \label{s3}

The persistent homology algorithm relies on building Vietoris-Rips complexes, which is computationally expensive when the dataset is large enough, and exacerbated by the need to build one complex for every $l$. While a typical EHT reconstructed image has only $\sim 80\times80$ pixels, the latest EHT imaging surveys generate millions of them. Therefore, a special preprocessing technique is required to reduce the number of datapoints $N$ in order to make this TDA algorithm practical for imaging survey at EHT scale.

Here, we introduce ``metronization'' and demonstrate that it is an effective preprocessing technique for persistent homology.
Metronization simplifies a VLBI image into a representation with less than 100 datapoints that still contain the same topological information of the original image.  This leads to over a thousand time speed up in the application of the TDA algorithm.

A secondary goal of metronization is to convert a black hole image, which is a three-dimensional data cloud in the space $(x, y, I)$, where $x$ and $y$ are the horizontal and vertical pixels (or angles) inscribed in the sky and $I$ the brightness, into a series of two-dimensional data clouds each in the space $(x,y)$ for every level of brightness. After metronization, emission rings will appear as circles in the two dimensional pixel space. While two-dimensional holes in the three-dimensional space $(x,y,I)$ or other high dimensional topological characteristics can contain important information on the image, we relegate its exploration to a future study.  

\begin{figure*}[h]
    \centering
    \includegraphics[width=6in]{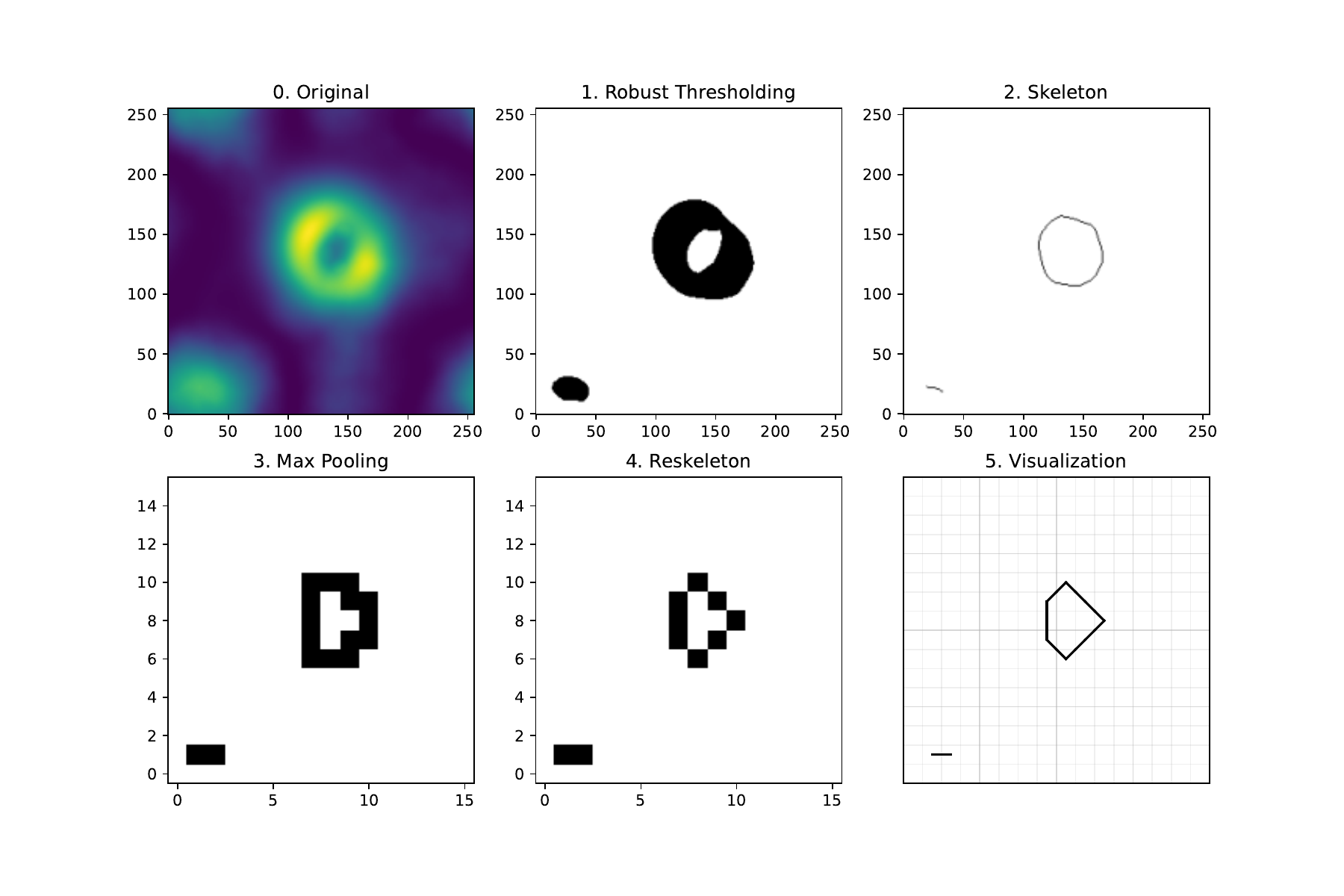}
    \caption{Example of a synthetic image undergoing the metronization procedure. The robust thresholding parameter is $T=0.7$. An emission ring can clearly be seen in the original image. While the metronization procedure wiped out much of the details in the original image, this topological feature is preserved.}
    \label{fig:sample1}
\end{figure*}

The metronization algorithm chains up different image processing methods to highlight the topological features in VLBI images. The steps involved in metronization preserve the topological characteristics of the original image, and thus the speed up obtained through metronization does not come at any cost in the accuracy of the TDA results. The different stages are shown in Figure~\ref{fig:sample1}.
The original image is shown in the top-left panel and the different steps are shown in the subsequent panels:
\begin{enumerate}
\item \emph{Robust thresholding.}  We first need to turn an image with scalar value pixels to boolean value pixels.  Instead of applying user-supplied threshold based on the peak intensity, we perform the following method.  First, we sort the pixel values $I(x_i, y_j)$ into a monotonic increasing sequence $(I_1, I_2, \dots, I_{N^2})$.  We then compute the cumulative sum of this sequence $(C_1, C_2, \dots, C_{N^2})$ such that $C_i = \sum_{i'\leq i} I_{i'}$.  It is clear that $C_i$ is also a monotonically increasing sequence.  We then apply a user supplied threshold $0 \leq T \leq 1$ by selecting the pixels corresponding to the indices $i$ such that $T \leq C_i/C_{N^2}$.  This step converts the black hole image from a three-dimensional data cloud in $(x,y,I)$ into a series of two-dimensional data clouds in $(x,y)$ for every $T$. Compared to thresholding based on the peak intensity, this method can be seen as an integration over pixel values weighted by their appearance frequency, and is much more robust against to the noise in the image.
\item \emph{Skeleton.}  We then apply a standard topological skeleton algorithm to the thresholded image at its native resolution.  Large area of flagged pixels will be thinned in this step, and mid-sized ``holes'' in the image will be enlarged.  Without this step, ``holes'' smaller than the max pooling resolution (see next) will be removed at the end. 
\item \emph{Max pooling.} Next, we down sample the image to, e.g., $16\times16$, pixels using max pooling.  That is, if any of the pixels in the rebinned pixel is flagged, we flag that pixel.
\item \emph{Reskeleton.} We then apply the skeleton algorithm again to the down sampled image.  This step removes unnecessary corner pixels.  The result is our metronized datapoints.
\item \emph{Visualization.} Finally, we visually connect the flagged pixels like in a metro map.  This step is for visualization only and is irrelevant to the application of persistent homology.
\end{enumerate}

\section{Classifying Synthetic Images} \label{s4}

We apply TDA on mock data by computing the persistent homology of the synthetic images. Our goal is to classify images which show emission ring features. We present two methods to convert the original image to a form suitable for TDA, the \emph{grid-scan} method and the \emph{bisection} method. 

\subsection{Grid-scan method}

In the \emph{grid-scan} method, we scanned through a set of robust thresholding parameters of $T=(0.1, 0.2, \ldots, 0.9)$, thus generating a grid in $T$-space. Higher grid resolution will result in a finer classification scheme at the cost of more computational time. In our case, nine instances of $b_1$ barcode diagrams are generated per image. The persistence parameter is set to be $L=0.5$, but as we will show in the next subsection, our result is not sensitive to changes in this parameter due to the large sizes of typical emission rings. 

If a persistent hole is detected at any of the nine $b_1$ diagrams, we classify the image as having a hole. Both $l$ and $T$ where the hole is detected is also recorded. The parameter $l$ gives an approximate size of the hole in the emission while the parameter $T$ encodes its brightness. The set $(l,T)$ can then be used to further classify the images in terms of the size and brightness of their emission rings. 

Figure \ref{fig:sample1} shows an example synthetic image being preprocessed by metronization with $T=0.7$. At $T=0.9$, the two bright blobs attached on the ring dominate, giving the image a zeroth Betti number of $b_0=2$ and a first Betti number of $b_1=0$. As $T$ is lowered, the two blobs meld into a single connected component and a persistent hole can now be detected. Consequently, the Betti numbers adjust to $b_0=1$ and $b_1=1$ to reflect this qualitative change. As $T$ is lowered further, the emission regions at the corners of the image start registering as separate connected components, and a corresponding increase in the zeroth Betti number $b_0$ is observed by the TDA algorithm. The changes in the number of connected components as $T$ is varied can be used to detect qualitative features such as bright emission blobs or completely disconnected emission regions, as demonstrated in this example.

\begin{figure*}[t]
    \centering
    \includegraphics[width=2.3in]{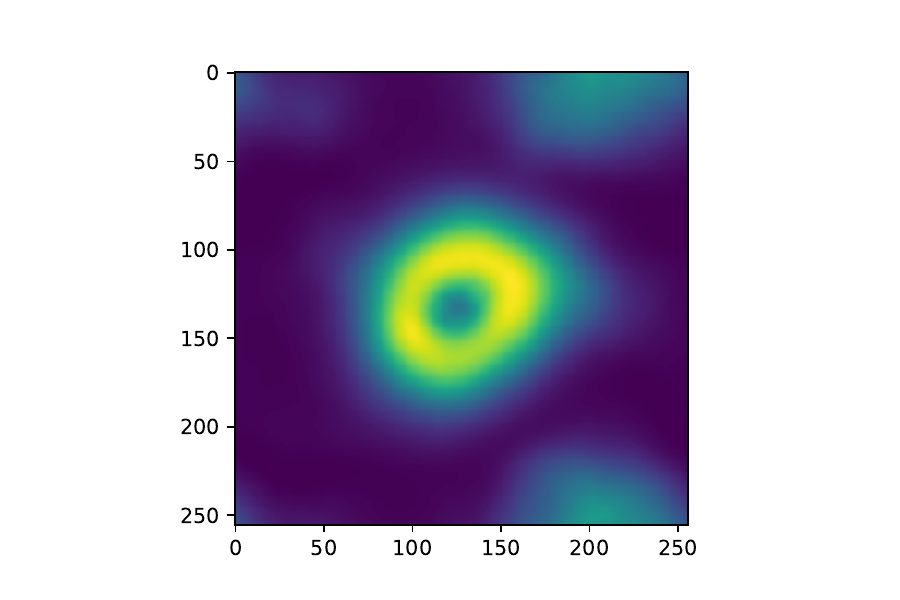} 
    \includegraphics[width=2.3in]{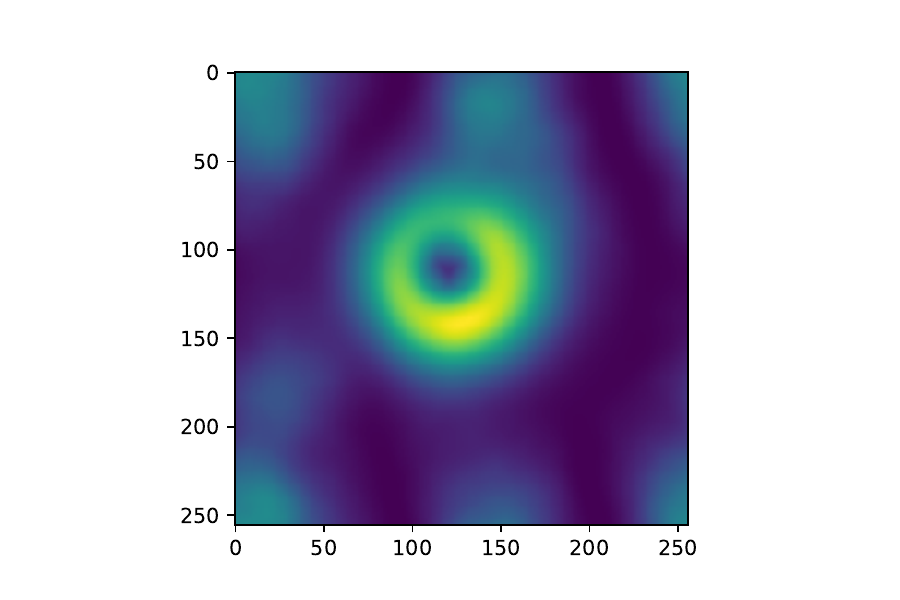}
    \includegraphics[width=2.3in]{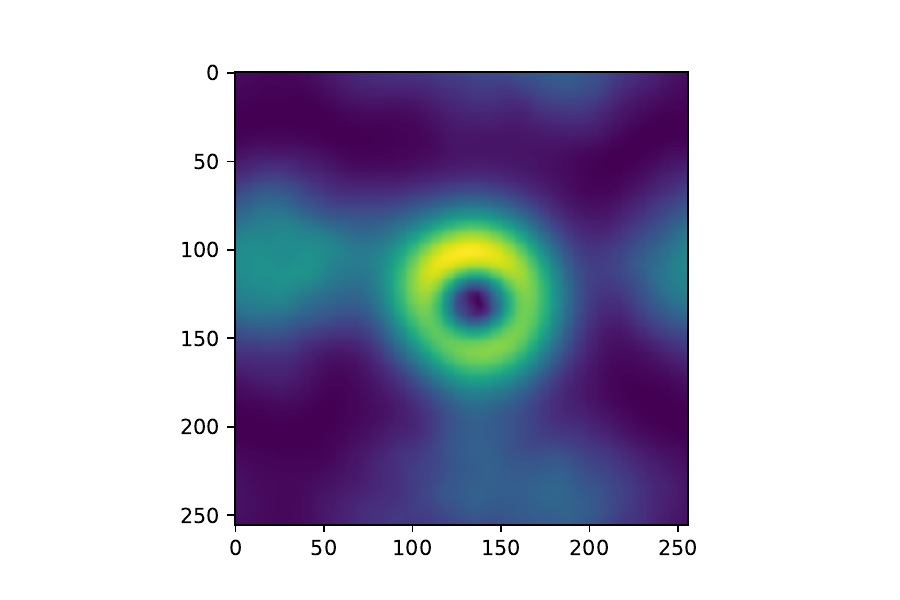} \\
    \includegraphics[width=2.3in]{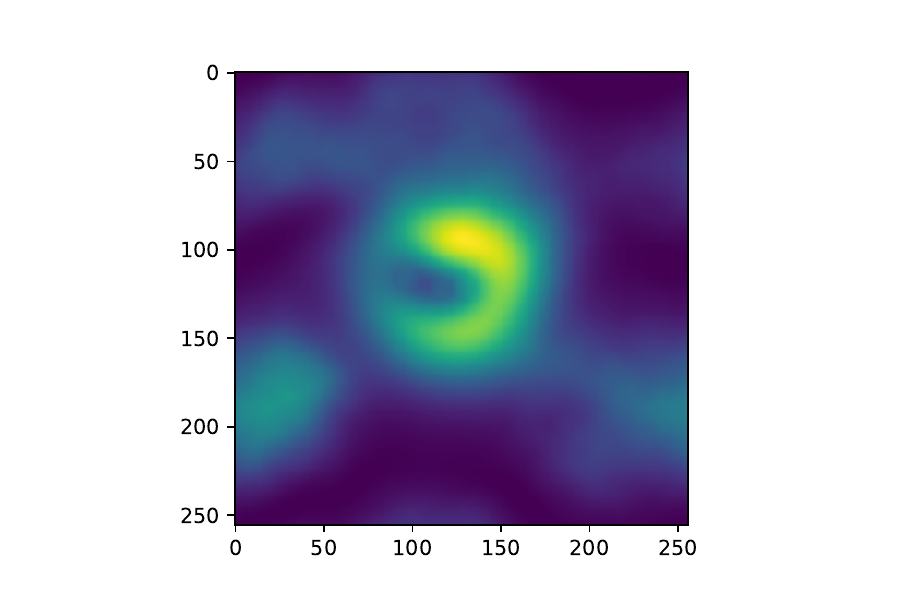}
    \includegraphics[width=2.3in]{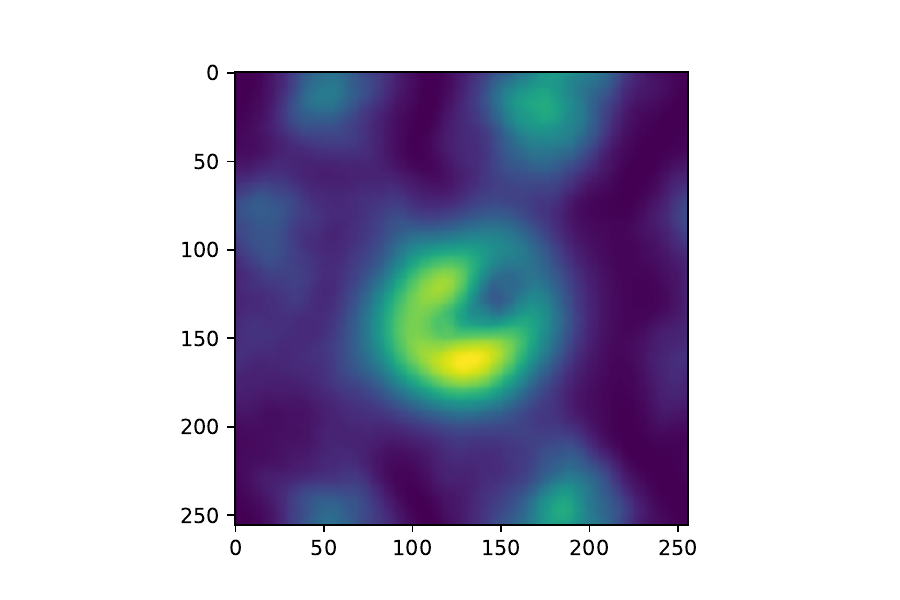}
    \includegraphics[width=2.3in]{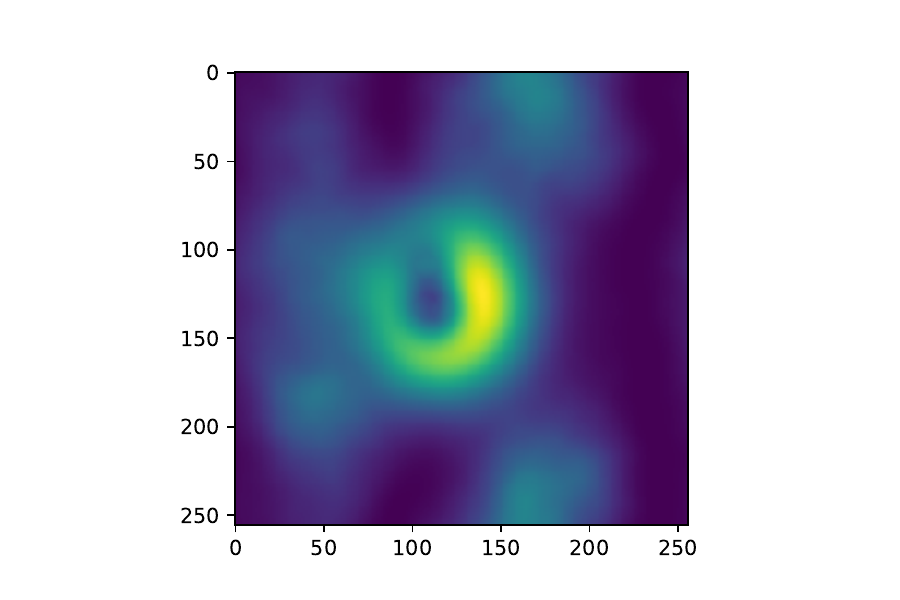}
    \caption{Examples of images classified as strong rings (top row) and images classified as weak rings (bottom row). From left to right, the strongly detected images posses persistent holes first detected at thresholds $T$ of $0.82$, $0.77$, and $0.84$, respectively. Similarly, the weakly detected images posses holes first detected at thresholds of $0.53$, $0.52$, and $0.64$ respectively.}
    \label{fig:more_examples}
\end{figure*}

Further sub-classifications can be made to characterize the images. For example, the $T$ where a persistent hole is first detected (when scanning from large to low $T$) can be used to characterize the likelihood of the hole's detection, as a persistent hole detected at high $T$ is less likely to be spontaneously created by noise. One can classify all images with holes detected at $T$ greater than a cutoff value as images with "strongly detected" holes, and those with holes detected under the cutoff to be ``weakly detected'' holes. In Figure \ref{fig:more_examples}, we show three images with ``strongly detected'' holes, which in this case means possessing a persistent hole that was detected at $T>0.75$ as well as three ``weakly detected'' holes. 

\subsection{Bisection method}

\begin{figure}[t]
    \centering
    \includegraphics[width=\linewidth]{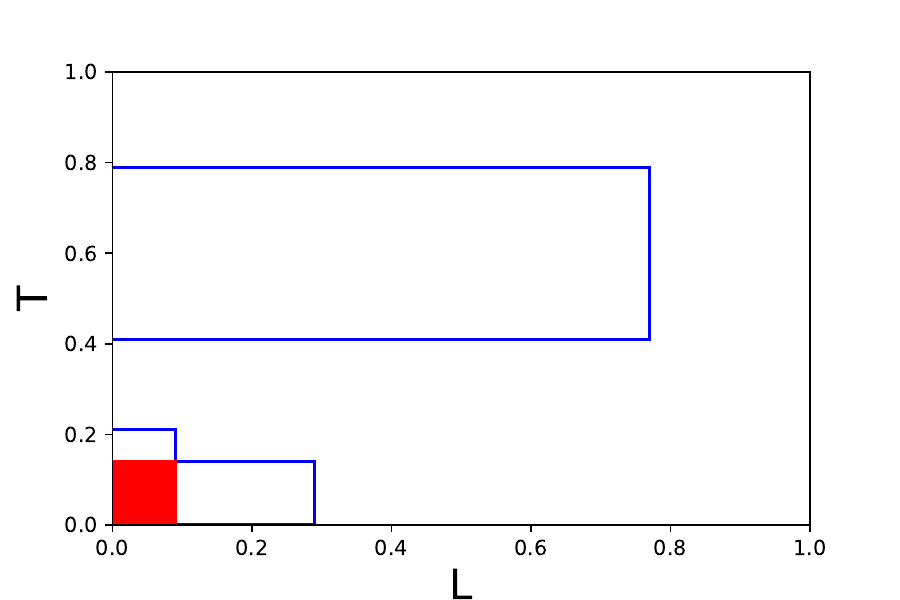} \\
    \caption{A "phase diagram" of the number of holes in $L-T$ space. Regions bordered in blue contain one hole, and regions filled in red contain four holes. Everywhere else zero holes are detected.}
    \label{fig:phase_diagram}
\end{figure}

As exemplified by Figure \ref{fig:sample1} in the previous subsection, changes in the Betti numbers can carry elucidating information that can be used to classify the image. In the \emph{bisection} method, we focus on locating the $T$ where changes in the Betti number is detected. 

Our bisection method took inspiration from the bisection method from root-finding, with a slight modification due to $b_i(T)$, the $i$th Betti number as a function of $T$, not being a monotonic function: 
\begin{enumerate}
    \item Start with a uniform grid in $T$-space, as in the grid-method, each of size $\mathrm{d} T$
    \item Assume that $b_i(T)$ is monotonic within a grid
    \item Suppose a Betti number changes between $T=T_\mathrm{high}$ and $T=T_\mathrm{low}$ (with $T_\mathrm{high}-T_\mathrm{low}<\mathrm{d} T$)
    \item Bisect the interval between $T_\mathrm{high}$ and $T_\mathrm{low}$ in half, and choose the interval where the Betti number changes for the next step
    \item Repeat the first step with the new interval
\end{enumerate}
The free parameter $\mathrm{d} T$ controls how fine the classification scheme is: the algorithm will not detect changes in $b_i(T)$ over intervals smaller than $\mathrm{d} T$. Running the bisection method with $\mathrm{d} T = 0.01$ allow us to determine borders in parameter space where different number of holes are detected in the image. When plotted in $L-T$ space (see Figure \ref{fig:phase_diagram}.), this gives a ``phase diagram'', indicating regions where different number of holes are detected. 

As seen in Figure \ref{fig:phase_diagram}., for a large swath of parameter space, one hole is detected. When the brightness threshold is too large (above $0.8$), dim pixels at the center of the emission ring started to contaminate the metronization result, filling the hole and thus resulting in zero holes being detected. As $L$ increases, we lose sensitivities to holes with radius smaller than $\sim 1.73 L$. Thus, when $L$ is too large, the emission ring is no longer detected. At the low $L$ and $T$ regime, noise dominates, resulting in multiple detected rings in the lower left region of Figure \ref{fig:phase_diagram}.

\subsection{Other Parameters}

While this work focuses on the detection of the emission rings themselves, we note that parameters other than $L$ and $T$ can be used to further classify black hole images or input a priori information into the classification system. Important examples are the characteristic lengths where the barcode diagram begins $l_{\rm birth}$ and ends $l_{\rm death}$, the analog of lifespan in $T$-space, as well as the zeroth Betti number, $b_0$.

The parameter $l_\mathrm{birth}$ determines the largest gap in a possible ring that could be tolerated (see Figure \ref{fig:birthdeathschem}). Any persistent holes that are detected at large $l_{\rm birth}$ will have a large gap (or gaps) in its surrounding emission ring. If $l_{\rm birth}$ is set to be very high, crescents will be detected as holes in the $b_1$ diagram. A strategy to classify crescents in the images is therefore to find images with $b_1=1$ (a hole is detected) and $b_0=1$ (a single piece is present), but with large $l_{\rm birth}$'s. Analogously, one can demand a maximum $l_{\rm birth}$ for an image to be counted as possessing a persistent hole in addition to the requirement that $\Delta l > L$ in order to remove crescent images from the population. 

As mentioned in the previous sections, the parameter $l_\mathrm{death}$ contains information about the size of the detected hole. A data cloud with holes that are small compared to $l_\mathrm{death}$ (i.e., with radii of less than $\approx 1.73 l_\mathrm{death}$ for circular holes) will have the holes filled in when converted into simplicial complex. Thus, the parameter $l_\mathrm{death}$ can be used to classify holes based on their sizes. Setting a minimum $l_\mathrm{death}$ filters holes that are too small, while setting a maximum $l_\mathrm{death}$ filters holes that are too big. If there is a priori information on the possible sizes of the hole in the image (e.g., from an independent measurement of the black hole mass), then both a minimum and maximum $l_{\rm death}$ can be used to cull images possessing holes that are inconsistent with the a priori possible hole size. 

A third parameter would be the analog of lifespan in $T$-space. For an image like that of Figure \ref{fig:sample1}, there will be no hole detected at high brightness thresholds, as the bright spots dominate the image. As $T$ is lowered, the emission ring becomes dominant, and a hole is detected. However, as $T$ is lowered further, the brightness depression at the center of the emission ring becomes incorporated into the simplicial complex, filling the hole. There is thus a threshold $T=T_{\rm max}$ where the hole first becomes detected, and another threshold $T_{\rm min}$ where the hole is no longer detected, and a $\Delta T = T_{\rm max} - T_{\rm min}$. For emission rings, $\Delta T$ is directly related to the brightness contrast between the emission ring and the brightness depression of its center. Images with emission rings with a large contrast will possess high $\Delta T$. 

Further, the zeroth Betti number, $b_0$, which counts the number of connected components in the image can also be used as a proxy of how noisy the image is, as a noisy image will possess a large number of connected pieces that are disconnected from each other. The images can then be sub-classified to be either ``noisy" or ``clean" images based on the number of, for example, the disconnected components larger than a certain size, detected at a certain threshold $T$. The first Betti number, $b_1$ can also be used for this purpose: noisy images will have more spurious holes detected at low $L$ and $T$ than clean images.

\section{Conclusion} \label{s5}

In the near future, both image reconstructions from VLBI dataset (such as that obtained by the EHT) and synthetic observations generated from theoretical models are poised to generate a very large library of black hole images. In this article we proposed that persistent homology aided with metronization pre-processing is a useful tool to automatically classify these images based on their topological characteristics. This method is both computationally efficient and flexible, in the sense of it possessing free parameters that could be tweaked based on the needs of the problem under consideration.

In this article we focus on the detection of emission rings based on the first persistent Betti number, $b_1$, and demonstrate how TDA can be used to not only separate images with rings and without rings, but also sub-classify ringed images into "weak" or "strong" rings based on their detection threshold. We also discussed ways to further leverage TDA to classify black hole images based on the sizes and brightness contrast of their emission rings, how crescent-like the image is, as well as the noise level of the image. 

Black hole images can contain astrophysically significant features aside from rings which will also imprint topological information on the image, and our method has the potential to classify images with these features. Images with hotspots will have $b_0>1$ at high $T$'s, with the number of distinct connected components giving the number of hotspots and the analog to $l_{\rm death}$ for the $b_0$ barcode diagram giving a measure for the distance between hotspots on the image plane. Simulations of black hole accretion flows have also been shown to display a bright emission ring surrounding an ``inner shadow'' \citep{2021ApJ...918....6C}, giving a hole-within-a-ring structure to the image. Future high resolution VLBI observations have the potential to observe this structure. Such structure will generate images with two rings that can be detected with TDA, possibly at different brightness thresholds, and with different contrasts. 

\section*{Acknowledgements}
The authors thank Shiro Ikeda for serving as the EHT collaboration internal referee for this paper.

\bibliography{tdabib.bib}

\end{document}